\documentstyle[epsf]{elsart}

\newcommand{\krig}[1]{\stackrel{\circ}{#1}}

\begin{document}

\hfill TK 95 21

\bigskip 

\begin{frontmatter}
\title{BARYON MASSES AND PION--NUCLEON $\sigma$--TERM\\
       TO SECOND ORDER IN THE QUARK MASSES\thanksref{dfg}}
\thanks[dfg]{Work supported in part by the Deutsche Forschungsgemeinschaft}


\author[Bonn]{B.~Borasoy}\footnote{email: borasoy@pythia.itkp.uni-bonn.de},
\author[Bonn]{Ulf-G.~Mei\ss ner}\footnote{email:
                                          meissner@pythia.itkp.uni-bonn.de}


\address[Bonn]{Institut f\"ur Theoretische Kernphysik, Universit\"at
  Bonn\\ Nu\ss allee 14--16, D-53115 Bonn,  Germany}


\begin{abstract}
We analyze the octet baryon masses and the pion--nucleon
$\sigma$--term in the framework of heavy baryon chiral
perturbation theory. In contrast to previous investigations, we
include {\it all} terms up-to-and-including quadratic order in the
light quark masses. The pertinent low--energy constants are
fixed from resonance exchange. This leaves as the only free parameter
the baryon mass in the chiral limit, $\krig{m}$. We find $\krig{m} =
749 \pm 125$~MeV together with $\sigma_{\pi N} (0) = 48 \pm 10$ ~MeV.
We discuss various implications of these results.
\end{abstract}

\end{frontmatter}


\section{Introduction}
The analysis of the octet baryon masses in the framework of chiral
perturbation theory  already has a long history, see e.g.
\cite{lang,pagels,juerg,gl82,liz,dobo,bkmz,lelu,samir,mary,dent,buergi}.
In this paper, we present
the results of a first calculation including {\it all} terms of order
${\cal O}(m_q^2)$, where $m_q$ is a generic symbol for any one of the light
quark masses $m_{u,d,s}$. We work in the isospin limit $m_u = m_d$ and
neglect the electromagnetic corrections. Previous investigations only
considered mostly the so--called computable corrections of order
$m_q^2$ or included some of the finite terms at this order.
This, however, contradicts the spirit of chiral perturbation
theory (CHPT) in that all terms at a given order have to be retained, see
e.g. \cite{wein79,gl84,gl85}. In general, the quark mass expansion of
the octet baryon masses takes the form
\begin{equation}
m = \, \, \krig{m} + \sum_q \, B_q \, m_q + \sum_q \, C_q \, m_q^{3/2} +
\sum_q \, D_q \, m_q^2  + \ldots
\label{massform}
\end{equation}
modulo logs. Here, $\krig{m}$ is the mass in the chiral limit of
vanishing quark masses and the coefficients $B_q, C_q, D_q$ are
state--dependent. Furthermore, they include contributions proprotional
to some low--energy constants which appear beyond leading order in the
effective Lagrangian. In this letter, we  evaluate these
coefficients for the octet baryons $N, \Lambda, \Sigma$ and $\Xi$.
In addition, we also calculate the pion--nucleon $\sigma$--term which
is intimately related to the quark mass expansion of the baryon masses
\cite{juerg,bkmz,jms}. For some comprehensive reviews, see e.g.
\cite{ulfrev,ecker,bkmrev,toni}.


\section{Effective Lagrangian}

To perform the calculations, we make use of the effective
meson--baryon Lagrangian. Our notation is identical to the one
used in \cite{bkmz} and we discuss here only the new terms.
Denoting by $\phi$ the pseudoscalar Goldstone fields ($\pi, K, \eta$) and
by $B$ the baryon octet, the effective Lagrangian takes the form
\begin{equation}
{\cal L}_{\rm eff} = {\cal L}_{\phi B}^{(1)} +  {\cal L}_{\phi B}^{(2)} +
  {\cal L}_{\phi B}^{(4)} + {\cal L}_{\phi}^{(2)}+ {\cal L}_{\phi}^{(4)}
\label{leff}
\end{equation}
where the chiral dimension $(i)$ counts the number of derivatives
and/or meson mass insertions. The baryons are treated in the extreme
non--relativistic limit \cite{jm,bkkm}, i.e. they are characterized
by a four--velocity $v_\mu$. In this approach, there is a one--to--one
correspondence between the expansion in small momenta and quark masses
and the expansion in Goldstone boson loops, i.e. a consistent power
counting scheme emerges. The form of the lowest order meson--baryon
Lagrangian is standard, see e.g. \cite{bkmz}, and the
meson Lagrangian is given in \cite{gl85}.
The dimension two meson--baryon Lagrangian can be written as
(we only enumerate the terms which contribute)
\begin{equation}
{\cal L}_{\phi B}^{(2)} = {\cal L}_{\phi B}^{(2, {\rm stand})} +
\sum_{i=1}^{10} \, b_i \, O_i^{(2)} \, \, ,
\label{leff2}
\end{equation}
with the $O_i^{(2)}$ monomials in the fields of chiral dimension two. Typical
forms are ${\rm Tr}(\bar{B} [ u_\mu , [ u^\mu,B]])$,
${\rm Tr}(\bar{B} [ v \cdot u , [ v \cdot u,B]])$
or $\bar{B} B {\rm Tr} (u_\mu u^\mu ) $, with $u_\mu = i
u^\dagger \partial_\mu U u^\dagger$, $u = \sqrt{U}$ and $U=\exp(i \phi
/ F_P)$ collects the pseudoscalars. The form
of ${\cal L}_{\phi B}^{(2, {\rm stand})}$  is \cite{bkmz},
\begin{equation}
{\cal L}_{\phi B}^{(2, {\rm stand})} = b_D \, {\rm Tr}(\bar B \lbrace
\chi_+ , B \rbrace ) + b_F \, {\rm Tr}(\bar B [\chi_+,B]) +
b_0 \, {\rm Tr}(\bar B
B) \, {\rm Tr}(\chi_+ ) \, \, ,
\label{leff2st}
\end{equation}
i.e. it contains three low--energy constants and $\chi_+ = u^\dagger
\chi u^\dagger + u \chi^\dagger u$ is proportional to the quark mass
matrix ${\cal M} ={\rm diag}(m_u,m_d,m_s)$ since $\chi = 2 B {\cal
  M}$. Here, $B = - \langle 0 | \bar{q} q | 0 \rangle / F_P^2$ and $F_P$ is
the pseudoscalar decay constant.
All low--energy constants in ${\cal L}_{\phi B}^{(2)}$ are finite.
A subset of the $b_i$ has been estimated in \cite{norb} by
analyzing kaon--nucleon scattering data.
The splitting of the dimension two meson--baryon Lagrangian in
Eq.(\ref{leff2}) is motivated by the fact that while the first three
terms appear in tree and loop graphs, the latter ten only come in via
loops. There are seven terms contributing at dimension four,
\begin{equation}
{\cal L}_{\phi B}^{(4)} =
\sum_{i=1}^{7} \, d_i \, O_i^{(4)} \, \, ,
\label{leff4}
\end{equation}
with typical forms of the $O_i^{(4)}$ are $\bar{B}B {\rm Tr}
(\chi_+^2)$ or ${\rm Tr}(\bar{B}[\chi_+ , [\chi_+ , B]] )$.
At this stage,
we take $m_u = m_d \ne m_s$. For $m_u \ne m_d$, there is an additional
term in ${\cal L}_{\phi B}^{(4)}$. The
explicit expressions for the various terms in
Eqs.(\ref{leff2},\ref{leff4}) can be found in \cite{bora}.
We point out that there are 20 a priori unkown constants. In addition,
there are the $F$ and $D$ coupling constants (subject to the
constraint $F+D = g_A =1.25$)
from the lowest order Lagrangian
${\cal L}_{\phi B}^{(1)}$.
What we have to calculate are all one--loop
graphs with insertions from ${\cal L}_{\phi B}^{(1,2)}$  and  tree
graphs from ${\cal L}_{\phi B}^{(2,4)}$. We stress that we do not
include the spin--3/2 decuplet in the effective field theory
\cite{dobo}, but rather use these fields to estimate the pertinent
low--energy constants (resonance saturation principle). We therefore
strictly count in small quark masses and external momenta with no
recourse to large $N_c$ arguments.


\section{Baryon masses and pion--nucleon $\sigma$--term}

The form of the terms $\sim m_q $ and $\sim m_q^{3/2}$
for the baryon masses and $\sigma_{\pi N} (0)$ is standard, we
use here the same notation as Ref.\cite{bkmz}. The $q^4$
contribution to any octet baryon mass $m_B$ takes the form
\begin{eqnarray}
m_B^{(4)} & = & \epsilon_{1,B}^P \, M_P^4 +
\epsilon_{2,B}^{PQ} \, M_P^2 \, M_Q^2 \nonumber \\
\quad & \, + &
\epsilon_{3,B}^P \, M_P^4 \, \ln ( \frac{M_P^2}{\lambda^2}) +
\epsilon_{4,B}^{PQ} \, M_P^2 \, M_Q^2 \, \ln ( \frac{
  M_P^2}{\lambda^2} )
\, \, ,
\label{mB4}
\end{eqnarray}
with $P,Q = \pi , K, \eta$ and $\lambda$ the scale of dimensional
regularization. The explicit form of the state--dependent
prefactors $\epsilon_{i,B}$ can be found in Ref.\cite{bora}.
Notice the  appearance of mixed terms $\sim M_P^2 \, M_Q^2$
which were not considered in most existing investigations. The fourth
order contribution to the baryon masses contains divergences
proportional to (using dimensional regularization)
\begin{equation}
L = \frac{\lambda^{d-4}}{16 \pi^2} \biggl\lbrace \frac{1}{d-4} -
\frac{1}{2}[\ln (4 \pi) +1 - \gamma_E] \biggr\rbrace
\, \, ,
\label{L}
\end{equation}
with $\gamma_E = 0.572215$.
These are renormalized by an appropriate choice of the low--energy
constants $d_i$,
\begin{equation}
d_i = d_i^r (\lambda) + \Gamma_i \, L
\, \, .
\label{divd}
\end{equation}
In fact, all seven $d_i$  are divergent. The appearance
of these divergences is in marked contrast to the $q^3$ calculation
which is completely finite (in the heavy fermion approach). In what
follows, we set $\lambda =1$~GeV.
Similarly, the fourth order contribution to the pion--nucleon $\sigma$--term
can be written as
\begin{equation}
\sigma_{\pi N}^{(4)}(0)  =
M_\pi^2 \, \biggl[ \, \epsilon_{1,\sigma}^P \, M_P^2 +
\epsilon_{2,\sigma}^P \, M_P^2 \, \ln(\frac{M_P^2}{\lambda^2}) +
\epsilon_{3,\sigma}^{PQ} \, M_P^2 \, \ln(\frac{M_Q^2}{\lambda^2}) \, \biggr]
\, \, ,
\label{sigma4}
\end{equation}
with the $\epsilon_{i,\sigma}$ given in \cite{bora}.
Here, the renormalization is somewhat more tricky. It can most
efficiently performed in a basis of a linearly independent subset
of the operators $dO_i^{(4)}/ dm_q$, $q=(u,d,s)$, as detailed in
Ref.\cite{bora}. A good check on the rather lengthy expressions for
the nucleon mass and $\sigma_{\pi N} (0)$ is given by the
Feynman--Hellmann theorem, $\hat{m}(\partial m_N / \partial
\hat{m}) = \sigma_{\pi N} (0)$, with $\hat m$ the average light quark
mass, $\hat m = (m_u + m_d)/2$.
We remark here that in contrast to the order $q^3$
calculation, the shift to the Cheng--Dashen point, $\sigma_{\pi N}
(2M_\pi^2) - \sigma_{\pi N} (0)$, is no longer finite, i.e. there
appear $t$--dependent divergences. We therefore do not consider this
$\sigma$--term shift in what follows. We will also not discuss in
detail the two kaon--nucleon $\sigma$--terms,
$\sigma_{KN}^{(1,2)} (t)$, for similar reasons in this letter.


\section{Resonance saturation}

\begin{figure}[t]
\hskip 1.2in
\epsfysize=1.5in
\epsffile{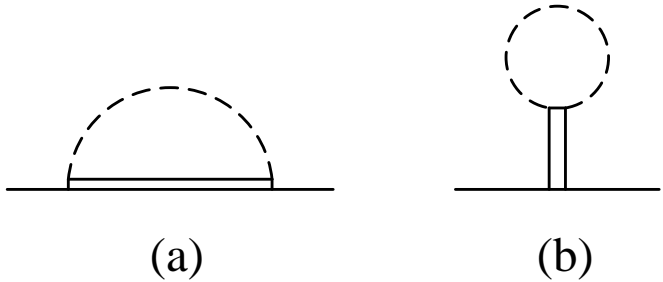}
\caption{\label{fig1} Resonance saturation for the masses. In (a), a
baryon resonance (from the decuplet) is excited, whereas (b) represents
the t--channel meson (scalar or vector) excitation. Solid and dashed
lines denote the octet baryons and Goldstones, respectively.}
\end{figure}

Clearly, we are not able to fix all the low--energy
constants appearing in ${\cal L}_{\phi B}^{(2,4)}$
from data, even if we would resort to large $N_c$ arguments.
We will therefore use the principle of resonance saturation to
estimate these constants. This works very accurately in the meson
sector \cite{reso,reso1,reso2}. In the baryon case, one has to account for
excitations of meson ($R$) and baryon ($N^*$) resonances. One writes
down the effective Lagrangian with these resonances chirally coupled to
the Goldstones and the baryon octet, calculates the pertinent Feynman
diagrams to the process under consideration and, finally,
lets the resonance masses go to infinity (with fixed ratios of
coupling constants to masses). This generates higher order terms in
the effective meson--baryon Lagrangian with coefficients expressed in
terms of a few known resonance parameters. Symbolically, we can write
\begin{equation}
\tilde{{\cal L}}_{\rm eff} [\, U,B,R,N^* \, ] \to
{\cal L}_{\rm eff} [\, U,B \, ]  \, \, .
\end{equation}
Here, there are two relevant contributions. One comes from the excitation
of the  spin-3/2 decuplet states and the second from t--channel scalar
and vector meson excitations, cf. Fig.\ref{fig1}. It is important to
stress that for the resonance contribution to the baryon masses, one
has to involve Goldstone boson loops. This is different from the
normal situation like e.g. in form factors or scattering processes.
Consider first the decuplet contribution. We treat these field
relativistically and only at the last stage let the mass become very
large. The pertinent interaction Lagrangian between the spin--3/2
fields (denoted by $\Delta$), the baryon octet and the Goldstones reads
(suppressing SU(3) indices)
\begin{equation}
{\cal L}_{\Delta B \phi} =
\frac{{\cal C}}{\sqrt{2} F_P} \, \bar{\Delta}^\nu \, T
  \, \Theta_{\nu \mu} (Z) \, B \, \partial^\mu \phi + {\rm h.c.} \, \, ,
\label{lmbd}
\end{equation}
with $T$ the standard $\frac{1}{2} \to \frac{3}{2}$ isospin transition
operator, $F_P = 100$~MeV the (average) pseudeoscalar decay constant
and ${\cal C} =1.5$ determined from the decays $\Delta \to B \pi$. The Dirac
matrix operator $\Theta_{\mu \nu} (Z)$ is given by
\begin{equation}
\Theta_{\mu \nu} (Z) = g_{\mu \nu} - \biggl(Z + \frac{1}{2} \biggr) \,
\gamma_\mu \, \gamma_\nu \, \, \, \, .
\label{theta}
\end{equation}
For the off--shell parameter $Z$, we use $Z = -0.3$ from the
determination of the $\Delta$ contribution to the $\pi N$ scattering
volume $a_{33}$ \cite{armin}.
The tadpole graphs shown in Fig.~1b with an intermediate vector meson
only contribute at order $q^5$. This is evident in the conventional
vector field formulation. In the tensor field formulation, the vertex
${\rm Tr}(V_{\mu \nu}u^\mu u^\nu)$ seems to lead to a contribution at
order $q^4$. However, in a tadpole graph one needs the index
contraction $\mu =\nu$ and thus has no term at order $q^4$. Matters
are different for the scalars. Denoting by $S$ and $S_1$ the scalar
octet and singlet with $M_S \simeq M_{S_1} \simeq 1$~GeV, respectively,
the lowest order coupling to the baryon octet reads
\begin{equation}
{\cal L}_{SB}^{(0)} = D_S \, {\rm Tr}(\bar B \lbrace S,B \rbrace )
+ F_S \, {\rm Tr}(\bar B [ S,B ] ) + D_{S_1} \,S_1 \, {\rm Tr}(\bar B B  )
\label{sb}
\end{equation}
where the coupling constants $D_S$, $F_S$ and $D_{S_1}$ are of the
order one.\footnote{In what follows, we neglect the singlet field.} In
fact, there are no empirical data to really pin them down. From the
decay pattern of the low--lying baryons one can, however, estimate
these numbers to be small. We will generously vary them between zero
and one. For the couplings of the scalars to the Goldstones, we use the
notation of \cite{reso} and the parameters determined therein.
Putting pieces together, all low--energy constants are expressed via
resonance parameters and the baryon masses take the form
\begin{eqnarray}
m_B & = & \krig{m} + m_B^{(3)}
+ \lambda_B \, ({\krig{m}})^{-1}
+ \beta \,  D_B^\beta
+ \delta \, D_B^\delta
+ \epsilon \, D_B^\epsilon + D_B^S
\quad , \nonumber \\
\beta & = & -\frac{1}{96 \pi^2} \, \frac{m_\Delta^4}{{\krig{m}}^3} \, , \quad
\delta= -\frac{\beta}{4{\krig{m}}^2} \, , \quad
\epsilon = \frac{2}{3m_\Delta} (2Z^2 + Z -1 )
\, \, \, ,
\label{massres}
\end{eqnarray}
with $ m_B^{(3)}$ the contribution of ${\cal O}(m_q^{3/2})$ and
$m_\Delta = 1.455$~GeV is the average decuplet mass.
Notice that it is convenient to lump the ${\cal O}(m_q)$ and
the ${\cal O}(m_q^2)$ corrections togther as it was done
in Eq.(\ref{massres}) (in the $D_B^\beta$ and $D_B^S$).
We have kept explict the baryon mass in the chiral limit. Of course,
in the fourth order terms it could be substituted by the corresponding
physical values. At second order, however, we would get a
state--dependent shift, see e.g. \cite{bkmzas}, and we thus prefer to
work with $\krig m$. Alternative representations
for the baryon masses are given in \cite{bora}.
Let us briefly explain the origin of the
various terms in Eq.(\ref{massres}). The $\lambda$ contributions are
tadpoles with $1/m$ insertions from ${\cal L}_{\phi B}^{(2)}$. Similarly,
the $\beta$ and $\epsilon$ terms stem from tadpole graphs with
insertions proportional to the low--energy constants $b_{0,D,F}$ and
$b_i$, respectively. Note, however, that in the resonance exchange
approximation not all of the ten $b_i$ are contributing. The
$\delta$ terms subsume the contributions from  ${\cal L}_{\phi B}^{(4)}$,
these are proportional to the low--energy constants $d_i$.
Finally, the terms of the type $D_B^S$ are the scalar meson
contributions to the mass. They amount to  a constant, state--dependent shift.
These consist of terms of the types $\sim M_P^2$, $\sim M_P^4 \, \ln
M_P^4$ and $\sim M_P^2 M_Q^2 \, \ln M_P^4$, see \cite{bora}.
To be specific,
we give the coefficients $D^{\beta,\delta,\epsilon}$ and $\lambda$
for the nucleon,
\begin{eqnarray}
& \lambda_N     & =
\frac{1}{8 \pi^2 F^2_P} \biggl\lbrace
-\frac{3}{32} (D+F)^2 M_\pi^4 \ln \frac{M_\pi^2}{\lambda^2}
- \frac{1}{96}(D - 3F)^2 M_\eta^4 \ln \frac{M_\eta^2}{\lambda^2}
\nonumber \\
& \, &
- \frac{1}{16}(D^2 - 2DF +3F^2) M_K^4 \ln \frac{M_K^2}{\lambda^2}
\biggr\rbrace \, \, , \nonumber \\
& D_N^\beta     & = D_N^{\beta,(2)} + D_N^{\beta,(4)} =
\frac{{\cal C}^2 }{16 \pi^2 F^4_P} \biggl\lbrace
(-4\pi^2 F_P^2)(M_K^2 + 4M_\pi^2) \nonumber \\
& + & \frac{27}{16}  M_\pi^4 \ln \frac{M_\pi^2}{\lambda^2}
+\frac{33}{24} M_K^4 \ln \frac{M_K^2}{\lambda^2}
+ \frac{7}{12} M_\eta^4 \ln \frac{M_\eta^2}{\lambda^2} \nonumber \\
& + & \biggl[ -\frac{11}{18}M_K^2 + \frac{43}{144}M_\pi^2
+\frac{1}{8}(D - 3F)^2 \bigl( \frac{M_K^2}{4}+M_\pi^2 \bigr) \biggr]
M_\eta^2 \ln \frac{M_\eta^2}{\lambda^2} \nonumber \\
 & + & \biggl[
\frac{9}{8}(D + F)^2 \bigl( \frac{M_K^2}{4} +M_\pi^2 \bigr) \biggr]
M_\pi^2 \ln \frac{M_\pi^2}{\lambda^2} \nonumber \\
 & + & \biggl[
-\frac{3}{4}(D - F)^2 M_K^2 + \frac{1}{96} (223D^2 - 318DF +351F^2) M_\pi^2
\biggr] M_K^2 \ln \frac{M_K^2}{\lambda^2}
\biggr\rbrace \, \, , \nonumber \\
& D_N^\delta    & =  \frac{{\cal C}^2}{2F_P^2} \biggl\lbrace
-\frac{415}{288} M_\pi^4 + \frac{83}{72} M_\pi^2 M_K^2 -
\frac{31}{72} M_K^4 \biggr\rbrace \, \, , \nonumber \\
& D_N^\epsilon  & =  \frac{{\cal C}^2}{8 \pi^2 F_P^2} \biggl\lbrace
- \frac{1}{2} M_\pi^4 \ln \frac{M_\pi^2}{\lambda^2}
- \frac{1}{8} M_K^4 \ln \frac{M_K^2}{\lambda^2} \biggr\rbrace
\, \, .
\label{DN}
\end{eqnarray}
The corresponding coefficients for the $\Lambda$, $\Sigma$ and $\Xi$
and also the $D_B^S$
can be found in \cite{bora}. The tree contribution from
${\cal L}_{\phi B}^{(2, {\rm stand})}$ is subsumed in the $D_B^\beta$.
The numerical values of the $\lambda_B$, $D^\beta_B, D^\delta_B$,
$D^\epsilon_B$ and $D_B^S$ are given in table~1
(using $F = 0.5$, $D=0.75$ and $c_{d,m}$ from \cite{reso}).
\begin{table}[h]
  \begin{tabular}{|c|ccccc|}
    \hline
    B &  $\lambda_B$ [GeV$^{2}$]
      & $D^\beta_B$ & $D^\delta_B$  [GeV$^{2}$]
      & $D^\epsilon_B$  [GeV$^{2}$] & $D_B^S$ [GeV] \\
    \hline
    $N$        & 0.0049   & $-$39.875 & $-$2.3617 & 0.0321
               &  $-0.017 \, D_S + 0.061 \, F_S$ \\
    $\Lambda$  & 0.0179   & $-$68.044 & 1.0831  & 0.0617 
               &  $-0.047 \, D_S - 0.010 \, F_S$ \\
    $\Sigma $  & 0.0144   & $-$132.90 & $-$11.273 & 0.1422 
               &  $+0.051 \, D_S + 0.010 \, F_S$ \\
    $\Xi$      & 0.0216   & $-$126.46 & $-$5.4799 & 0.1324 
               &  $-0.037 \, D_S - 0.082 \, F_S$ \\
    \hline
  \end{tabular}
  \medskip
  \caption{Numerical values of the state--dependent  coefficients
           in Eq.(\protect\ref{massres}). The $D_B^\beta$ are
           dimensionless. The $D_B^S$ are for $c_d$, $c_m$ from
           \protect\cite{reso} and $M_S =1$ GeV.}
\end{table}
We see that the dominant terms at ${\cal O}(m_q^2)$ are indeed the
tadpole graphs with an insertion from ${\cal L}_{\phi B}^{(2, {\rm
    stand})}$ (this holds for the masses but not for $\sigma_{\pi N}(0)$).
It is also instructive to compare the values we find from resonance
exchange with the ones previously determined from KN scattering
data. We have transformed the results of
Ref.\cite{norb} into our notation.   As can be seen from
table~2, most (but not all) coefficients agree in sign and magnitude.
\begin{table}[bht]
  \begin{tabular}{|c|ccccccc|}
    \hline
    & $b_1$ & $b_2$ & $b_3$ & $b_8$  & $b_0$ & $b_F$ & $b_D$ \\
    \hline
    Reso. $(D_S=F_S=0)$  & 0.084 & $-$0.144 & 0.108 & $-$0.216 & $-$0.738
                         & $-$0.264 & 0.317 \\
    Reso. $(D_S=F_S=1)$  & 0.100 & $-$0.111 & 0.125 & $-$0.239 & $-$0.733
                         & $-$0.208 & 0.345 \\
    Ref.\cite{norb} & 0.044 & $-$0.145 & $-$0.054 & $-$0.165 & $-$0.493
                    & $-$0.213 & 0.066 \\
    \hline
  \end{tabular}
  \medskip
  \caption{Some low--energy constants from ${\cal L}_{\phi B}^{(2)}$
            in GeV$^{-1}$. In the first row, only the decuplet
            contribution is given. In the second row, scalar meson
            exchange is added.}
\end{table}
Note, however, that this is only a subset of the coefficients
considered in this work. The full list will be given in \cite{bora}.
We remark that the procedure used in \cite{norb} involves the summation of
arbitrary high orders via a Lippmann--Schwinger
equation and is thus afflicted with
some uncertainty not controled within CHPT.


\section{Results and discussion}

The only free parameter in the formula for the baryon masses,
Eq.(\ref{massres}), is the mass of the baryons in the chiral limit,
$\krig m$, since all low--energy constants are fixed in terms of
resonance parameters. In particular, in contrast to Ref.\cite{bkmz},
the parameters $b_0$, $b_D$ and $b_F$ are no longer free. Also, at
quadratic order in the quark masses the ambiguity between $\krig m$
and $b_0$ is resolved, it is not  necessary to involve any one
of the $\sigma$--terms in the fitting procedure \cite{liz,bkmz}.
In fact, one can not find one single value
of $\krig m$ to fit all four octet masses, $m_N = 0.9389$~GeV, $m_\Lambda =
1.1156$~GeV, $m_\Sigma = 1.1931$~GeV and $m_\Xi = 1.3181$~GeV exactly. We
therefore fit these masses individually and average the corresponding
values for $\krig m$.
The contribution from scalar meson exchange only enters the
uncertainties of the numbers given. This is justified since the
numerical  values for the couplings $F_S$ and $D_S$ are supposedly small.
A more thorough discussion on this point can be found in \cite{bora}.
We have
$\krig{m}_N = 711 \, \, {\rm MeV}$,
$\krig{m}_\Lambda = 679 \, \, {\rm MeV}$,
$\krig{m}_\Sigma  = 877 \, \, {\rm MeV}$,
$\krig{m}_\Xi     = 728 \, \, {\rm MeV}$,
with the following average
\begin{equation}
\krig{m} = 749 \pm 125 \, \, {\rm MeV}\,\, . \quad
\label{mkrig}
\end{equation}
This number is compatible with the one found  in the analysis
of the pion--nucleon $\sigma$--term, where approximately 130 MeV to the
nucleon mass were attributed to the strange matrix element $m_s
\langle p | \bar s s|p \rangle$ (with a sizeable uncertainty) \cite{gls}.
The spread of the various values is a good measure of the
uncertainties related to this complete $q^4$ calculation.
Let us take a closer look at the quark mass expansion of the
nucleon mass, in the notation of Eq.(\ref{massform}),
\begin{equation}
m_N = 711 + 202 - 272 + 298 \, \, \, {\rm MeV}
= 939 \, \, {\rm MeV}     \, \, .
\label{mnuc}
\end{equation}
This looks similar for the other octet baryons.
We conclude that the quark mass corrections of order $m_q$,
$m_q^{3/2}$ and $m_q^2$ are all of the same size.\footnote{Note that
  for certain values of the scalar couplings, the fourth order term
  $m_N^{(4)}$ can be significantly smaller. However, the nucleon mass
  is much more sensitive to the scalar contribution than the other
  octet masses. }
 In Ref.\cite{jms},
it was argued that only the leading non--analytic corrections (LNAC)
$\sim m_q^{3/2}$ are large and that further terms like the ones $\sim m_q^2$
are modestly small, of the order of 100 MeV. This would amount to
an expansion in $M_K^2/(4 \pi F_P)^2 \sim 1/6$
with a large leading term. This expectation is not
borne out by our results, the next corrections are as large as the
LNACs. These findings agree with the meson cloud model calculation of
Gasser \cite{juerg}. A last remark about the baryon masses
concerns the deviation from the Gell-Mann-Okubo relation, $\Delta_{\rm
 GMO} = (3m_\Lambda+m_\Sigma-2m_N-2m_\Xi)/4$ which empirically is
about 6.5 MeV. We find  $\Delta_{\rm GMO} = 31$ MeV, which is larger
in magnitude than the value found in \cite{liz,dobo}. We remark
that in our case the decuplet contribution is contained in
the $m_q^2$ contributions and not in the $m_q^{3/2}$ as in \cite{liz}.
Therefore, in our case, $\Delta_{\rm GMO}$ is dominated by the
$m_q^2$ piece. The sizeable uncertainty in the chiral limit masses,
Eq.(\ref{mkrig}), does not allow for a very accurate statement about
this very small quantity. It is also very sensitive to the scalar couplings.
To get a better handle on this issue, one either has to be able to fix all
pertinent low--energy constants at order $m_q^2$ from data or improve upon the
resonance saturation estimate used here by including e.g. the mass splitting
within the decuplet and the SU(3) breaking of the
decuplet-octet-meson couplings. A better understanding of this topic
is, of course, at the heart of the determination of the quark mass
ratio $R = (m_d-m_u)/(m_s - \hat m )$ from the baryon masses (once the
electromagnetic corrections have been included).

The pion--nucleon $\sigma$--term is completely fixed. Using for
$\krig m$ its average, we find (no scalar resonance contribution)
\begin{equation}
\sigma_{\pi N} (0) = 48 \, \, {\rm MeV}     \, \, ,
\label{sigval}
\end{equation}
with an uncertainty of about $\pm 10$ MeV due to the spread in $\krig
m$. This number compares favourably  with the one found
in Ref.\cite{gls}, $\sigma_{\pi N} (0) = 44 \pm 9$~MeV. We stress that
this result reflects  a very non--trivial consistency
for the complete calculation to quadratic order in the quark masses
using the
resonance saturation principle in the scalar sector.
The additional contribution from the scalar meson exchange is
accounted for in the $\pm 10$ MeV uncertainty. To be specific, we have
$\sigma_{\pi N} (0) = (48 + D_S \cdot 0.5 - F_S \cdot 2.0) \, \,
{\rm MeV}$. It is furthermore
instructive to disentangle the various contributions to $\sigma_{\pi
  N} (0)$ of order $q^2$, $q^3$ and $q^4$, respectively,
\begin{equation}
\sigma_{\pi N} (0) = 54 - 33  + 27 \, \, \, {\rm MeV}
= 48 \, \, {\rm MeV}     \, \, ,
\label{signo}
\end{equation}
which shows a moderate convergence, i.e. the terms of increasing order
become successively smaller. Still, the $q^4$ contribution is
important. Also, at this order no $\pi \pi$ rescattering effects are
included. We notice that using the values for $b_{0,D,F}$ and $b_i$
as determined in Ref.\cite{norb} leads to a much increased fourth
order contribution.


\section{Summary and outlook}

In this paper, we have used heavy baryon chiral perturbation theory to
calculate the octet baryon masses to quadratic order in the quark
masses, including 20 local operators with unknown coeffcients. These
low--energy constants were fixed by resonance exchange.
The dominant contribution comes indeed from the excitation of the
spin--3/2 decuplet fields. Tadpole graphs with scalar meson exchange
only lead to small corrections. This left us
with one free parameter, the baryon mass in the chiral limit, which
could be determined within 18\% accuracy, and is compatible with the
value inferred for the strange matrix--element $m_s \langle p |  \bar
s s | p \rangle$ in Ref.\cite{gls}.
Furthermore, the pion--nucleon
$\sigma$--term  comes out surprisingly close to its empirical
value, $\sigma_{\pi N} (0) = 48$~MeV, with an uncertainty of
about $\pm 10$ MeV.
This first exploratory $q^4$ study of the three flavor scalar sector
of baryon CHPT points towards a significant improvement compared to
previous investigations which were mostly confined to so--called
``computable'' corrections and/or fitted a few of the pertinent
low--energy constants.
However, the calculation is not yet accurate enough to
determine the quark mass ratio $R$ reliably from the octet masses.
Furthermore, we did not address the kaon--nucleon $\sigma$--terms,
the corresponding shifts to the pertinent Cheng-Dashen points
together with the strangeness content of the proton here.
We will come back to these topics in Ref.\cite{bora}.


\section*{Acknowledgements}
We are grateful to Daniel Wyler for a very useful remark.


\end{document}